\documentclass[superscriptaddress,twocolumn,nofootinbib]{revtex4}
\usepackage{amssymb, bm}
\usepackage{graphicx}
\usepackage{hyperref}

\begin{document}

\title{Current-induced spin orientation in semiconductors and low-dimensional
structures}

\author{N.S. Averkiev}
\affiliation{A. F. Ioffe Physical-Technical Institute, Russian
Academy of Sciences, 194021 St. Petersburg, Russia}
\author{I.A. Kokurin}
\email{kokurinia@math.mrsu.ru} \affiliation{A. F. Ioffe
Physical-Technical Institute, Russian Academy of Sciences, 194021
St. Petersburg, Russia} \affiliation{Institute of Physics and
Chemistry, Mordovia State University, 430005 Saransk, Russia}

\date{\today}

\begin{abstract}
We present here a brief overview of current-induced spin
polarization in bulk semiconductors and semiconductor structures of
various dimension. The role of band structure and spin relaxation
processes is discussed. The related phenomena, such as spin Hall
effect, inverse spin Hall effect and other are discussed. Our recent
results in this field are presented as well.
\end{abstract}

\maketitle

Spin-orbit coupling (SOC) is a relativistic effect that provides a
link between spin and electric field (including the field of light
wave). The SOC is the basis of modern concept of semiconductor
spintronics. One can distinguish two main type of the spin
orientation at current carrying through the sample: (i) spin Hall
effect (SHE), which is the spatial separation of carriers with
opposite spins and (ii) homogeneous in the sample polarization. This
paper is focused mainly on effect of homogeneous current-induced
spin polarization (CISP), however, the related phenomena such as
SHE, inverse SHE and other are discussed as well. We will conduct
presentation adhering to the chronology of events, which coincides
with the transition from bulk semiconductors to low-dimensional
ones.

\begin{figure}[b]
\includegraphics[width=\columnwidth]{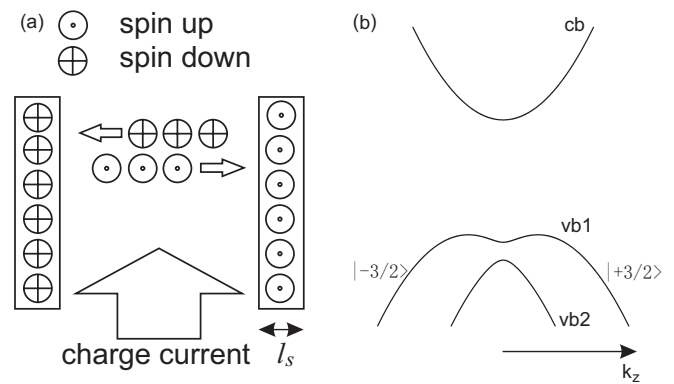}
\caption{\label{fig1} (a) Schematic description of SHE (top view).
The spins are accumulated in layers of width $l_s$ close to sample
edges. (b) The band structure of tellurium (schematically) near the
H-point of Brillouin zone. The valence bands are non-degenerate in
spin.}
\end{figure}

SHE is due to so-called Mott-scattering~\cite{Mott1965}, known in
atomic physics, that is the asymmetry of the scattering relative to
the plane determined by momentum and spin that in turn is due to
SOC. In semiconductors the role of SOC increases and such an effect
is several order of magnitude stronger. In 1971 Dyakonov and Perel
predicted this phenomenon in semiconductors~\cite{Dyakonov1971,
Dyakonov1971a}. The name SHE was introduced later by Hirsch in
1999~\cite{Hirsch1999} and the fist experimental observation of SHE
was in Awschalom group~\cite{Kato2004} through more than 30 years
after the prediction. The qualitative picture of SHE is depicted in
Fig.~\ref{fig1}a, where one can see, that the electrons with
opposite spin orientation experience the scattering predominantly in
opposite directions. This leads to accumulation of carriers with
opposite spin projection at opposite sides of the sample at the
length scales about spin diffusion length $l_s$.

The inverse SHE, i.e. the appearance of the dc current due to the
nonhomogeneous spin polarization, was predicted by Averkiev and
Dyakonov in 1983~\cite{Averkiev1983} and detected by Bakun et al.
\cite{Bakun1984} in the case, when inhomogeneous spin polarization
appears due to spin diffusion after interband excitation under
condition of optical orientation.

There are two main differences between CISP and SHE: (i) homogeneous
vs non-homogeneous spin polarization; (ii) CISP takes place only in
gyrotropic semiconductors, while SHE can be realized even in
structure with inversion center.

Phenomenologically CISP is described by second rank pseudotensor
$\hat{Q}$ linking the current density vector $\mathbf{j}$ and
average spin pseudovector $\mathbf{S}$

\begin{equation}
\label{phenomenology} S_i=\sum_kQ_{ik}j_k,\qquad i,k=x,y,z.
\end{equation}

From the symmetry point of view the gyrotropy means that the polar
vector (current density) and axial vector (spin) transform under the
same irreducible representation. Phenomenological description
(\ref{phenomenology}) shows that pseudotensor $\hat{Q}$ has non-zero
component $Q_{ik}$ only if $S_i$ and $j_k$ equally transform under
all symmetry operations of the point group of structure. Thus, the
relative orientation of spin $\mathbf{S}$ and current $\mathbf{j}$
is completely determined by the point group symmetry of the
structure. The details of symmetry analysis for frequent
two-dimensional (2D) structures one can find e.g. in
Ref.~\cite{Ganichev2014}.

In gyrotropic point symmetry groups there is no difference between
components of polar vector (e.g. electric field or current) and
axial vector (e.g. magnetic field or spin). The necessary
requirement for gyrotropy is the lack of inversion center. However,
there is misleading statement, that the gyrotropic crystalls have no
reflection planes. Nevertheless, this is too strict requirement, and
among 21 classes without spatial inversion only 3 ($T_d$, $C_{3h}$
and $D_{3h}$) are non-gyrotropic, and among the remaining 18 ones
only 11 are chiral (have no reflection planes or rotation-reflection
axes). The III-V-semiconductors with zinc-blende lattice possess
$T_d$-symmetry and hence are non-gyrotropic, but any symmetry
reduction such as strain or dimension lowering leads to CISP. The
wurtzite semiconductors (point group $C_{6v}$) are initially
gyrotropic.

For the first time, the possibility of CISP was proposed for bulk
semiconductor tellurium \cite{Ivchenko1978}. In this case due to
specific band structure (see Fig.~\ref{fig1}b), caused by strong
SOC, the electric current along trigonal axis leads to non-zero spin
oriented in that direction. The effect was detected by means of
additional rotation of polarization plane of the light
\cite{Vorobev1979}.

The carriers acquire the energy in electric field and this process
is counterbalanced by scattering, thus, electron distribution is
shifted to the following magnitude of quasi-momentum
\begin{equation}
\label{k_drift} \hbar {\bf k}_{dr}=e{\bf E}\tau.
\end{equation}
This anisotropy of carrier distribution leads to uncompensated
average spin in non-degenerate in spin band such as tellurium
valence band. Here ${\bf E}$ is the electric field and $\tau$ is the
momentum relaxation time.

In 2D electron structures the CISP mechanism is sufficiently
different from that in tellurium. The conduction band spin-splitting
is determined in 2D-structures by the Hamiltonian
\begin{equation}
\label{H_soc}
H=\frac{\hbar^2k^2}{2m^*}+\sum_{ij}\beta_{ij}\sigma_ik_j,
\end{equation}
where $m^*$ is the effective mass, $\sigma_i$ ($i=x,y,z$) are Pauli
matrices, $k_j$ ($j=x,y$) are the components of in-plane wave
vector, and components of pseudotensor $\hat{\beta}$ depend on
structure symmetry. From the symmetry point of view the tensors
$\hat{\beta}$ and $\hat{Q}$ in Eq.~(\ref{phenomenology}) are
equivalent.

The first mention on possibility of CISP in 2D-structure due to
Rashba spin-splitting~\cite{Bychkov1984} was in
Ref.~\cite{Vasko1979}. The consistent theory of CISP in
2D-structures was almost simultaneously proposed in
Refs.~\cite{Aronov1989,Edelstein1990}. In this connection CISP
frequently referred in literature as the Edelstein effect. More
complete theory for strained $\mathrm{A^{III}B^V}$ semiconductors
and 2D systems was developed in Ref. \cite{Aronov1991}. The
microscopical calculation of CISP is usually based on the solution
of quantum kinetic equation
\begin{equation}
\label{kinetic_eq} \frac{\partial\rho}{\partial t}
+\frac{i}{\hbar}[H^{so}_k,\rho]+\frac{e{\bf
E}}{\hbar}\frac{\partial\rho}{\partial {\bf k}}=\mathrm{St}\rho,
\end{equation}
where $\rho$ is the density matrix, that is diagonal in subband or
momentum index, $[A,B]$ stands for commutator, $H^{so}_k$ is the
Hamiltonian describing spin-orbit splitting of conduction band
(subband), $\mathrm{St}\rho$ is the collision integral taking into
account the processes of elastic scattering. This equation usually
can be solved by iterations taking into account the weakness of
spin-orbit splitting and electric field. It should be noted, that
SOC has to be taken into account in collision integral in contrast
to consideration of spin relaxation.

After~\cite{Aronov1989,Edelstein1990,Aronov1991} theoretical works
extensively reported about CISP in 2D-structures with different
symmetry and different types of
SOC~\cite{Chaplik2002,Averkiev2005,Raichev2007,Trushin2007,Golub2011},
such as structure inversion asymmetry (SIA)~\cite{Bychkov1984} due
to asymmetry of quantum well (QW) and bulk inversion asymmetry
(BIA)~\cite{Dyakonov1986} that is due to lack of inversion center in
semiconductor material~\cite{Dresselhaus1955}.

The average spin per particle can be estimated as
\begin{equation}
\label{spin} {\bf S}=r\frac{\hat{\beta}{\bf k}_{dr}}{\langle
E\rangle},\qquad r\sim 1,
\end{equation}
where $\langle E\rangle$ is the character energy, that is the
temperature in non-degenerate case and the Fermi energy $\mu$ in
degenerate one. A parameter $r$ depends on specific scattering
mechanism.

\begin{figure}
\includegraphics[width=\columnwidth]{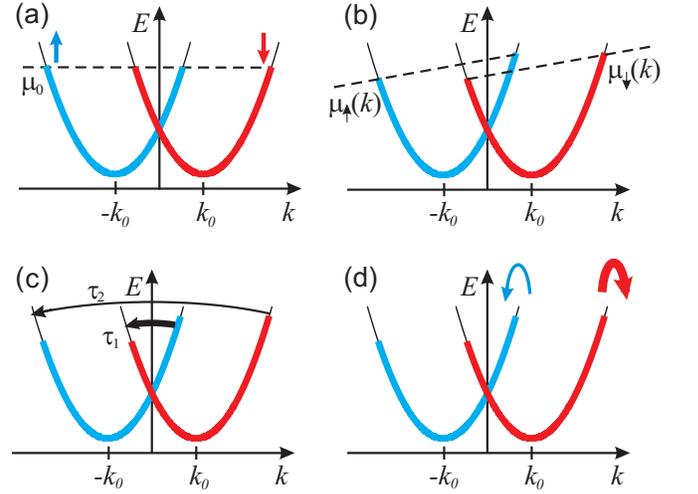}
\caption{\label{fig2} Qualitative explanation of CISP emergence in
2D structures. Three ingredients are required for CISP in 2D:
spin-orbit splitting (a); electric field, that alter
quasi-equilibrium carrier distribution (b); spin relaxation, that
can be both due to $k$-dependent relaxation time (c) and due to
$k$-dependence of precession frequency in DP-mechanism of spin
relaxation (d).}
\end{figure}

In 2D structures the microscopic mechanism of CISP is other than in
tellurium. The three ingredients are necessary for CISP in 2D-system
(see Fig.~\ref{fig2}): (i) spin-splitting; (ii) electric field, that
creates anisotropy of distribution function; (iii) spin relaxation
that can be due to $k$-dependent scattering or due to spin
precession by Dyakonov-Perel (DP) mechanism~\cite{Dyakonov1972}.

\begin{figure}
\includegraphics[width=\columnwidth]{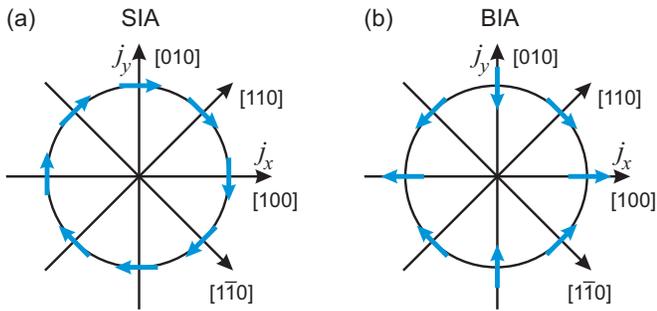}
\caption{\label{fig3} CISP in [001]-grown QW due to SIA (a) and BIA
(b).}
\end{figure}

CISP that arises in asymmetric QW due to SIA ($C_{\infty v}$ group)
is always perpendicular to the current direction (see
Fig.~\ref{fig3}a), while in the system with BIA ($D_{2d}$ group)
direction of spin polarization crucially depends on the current
direction relative to crystal axes as depicted in Fig.~\ref{fig3}b.
Thus, in [001]-grown QWs the charge current leads to in-plane spin
polarization only independently on symmetry of QW, its interfaces
and relative contribution of Rashba and Dresselhaus SOC-terms. The
magnitude of spin polarization in this case can be up to 5 percent.
The out-of-plane spin components can be generated in QW-structures
of lower symmetry, e.g. grown in direction [110], [112], [113] etc.
The exact relation may be established utilizing the symmetry
consideration only.

In significantly high electric field the regime known as 'streaming'
is realized, when the electron accelerates in electric field up to
the energy of optic phonon after that phonon is emitted and the
process will repeat. The CISP in this case can reach about 2\%
\cite{Golub2013}. Note here another mechanism of CISP: (i) based on
spin-dependent scattering even without
spin-splitting~\cite{Tarasenko2006}, (ii) bulk spin polarization
generated by the spin Hall current~\cite{Korenev2006}.

It seems, that CISP has to be more efficient in one-dimensional (1D)
systems (quantum wires) due to enhanced relaxation time according to
Eqs.~(\ref{spin}),(\ref{k_drift}). However, there is the theorem
asserting that in strictly 1D geometry all effects caused by SOC
(including CISP) are absent \cite{Entin2004}.

Nevertheless, we have recently shown \cite{Kokurin2015} that in
quasi-1D structure with more than one subband occupied due to
possibility of inter-subband scattering CISP is nonzero. For the
system with Rashba SOC the simple estimation was found in the case
of simple isotropic momentum relaxation
\begin{equation}
S_{1\rm D}\sim \frac{eE\Delta\alpha\tau_{1\rm D}}{\mu^2},
\end{equation}
where $\tau_{1\rm D}$ is the momentum relaxation time in 1D (for
simplicity we suppose the same time for intra- and inter-subband
scattering), and $\Delta$ is the energy between the ground and the
first excited quantized subband. This equation differs from that in
2D by replacement $\tau\rightarrow\tau_{1\rm D}\Delta/\mu$ [cf.
Eq.~(\ref{spin}) with $k_{dr}$ from (\ref{k_drift})]. Since
$\Delta/\mu<1$, but $\tau_{1\rm D}>\tau$ we can conclude that in
quasi-1D case the CISP has the efficiency close to that in 2D case.

Another picture can be realized in quasi-1D structures with strong
SOC. In this case the spin-splitted subbands with two minima and one
maximum (W-shaped) take place, e.g. in InAs-nanowires with tubular
electron gas~\cite{Kokurin2014,Kokurin2015a}, that resembles the
tellurium valence band spectrum. The main difference from tellurium
(where extrema of valence band are at side valleys) is the two-fold
Kramers degeneracy, because the electrons in mentioned structure are
at $\Gamma$-point of Brillouin zone. Nevertheless the applied
magnetic field lifts above degeneracy and CISP can be separated from
equilibrium polarization that is due to applied field. An advantage
of InAs-nanowires is the possibility to alter concentration (or
position of Fermi energy) by gate, that significantly affect the
CISP and average spin has non-monotonic dependence on Fermi energy.
Our estimations give close to 10\% depending on concentration at
reasonable magnitudes of electric field and structure parameters.

Let us now discuss some experimental techniques to measure CISP. In
tellurium the electric current $j_z$ applied along $C_3$-axis leads
to spin oriented in the same direction, $S_z\propto j_z$. In
experiment~\cite{Vorobev1979} the Faraday rotation was utilized to
detect the CISP-effect. However, tellurium possesses the natural
optical activity, and rotates the polarization plane of linearly
polarized light without any electric field or currents.
Nevertheless, because of the linear-in-current dependence the
CISP-effect was separated by means of inversion of current
direction.

In QWs the effect was registered using a circular polarization of
photoluminescence at additional non-polarized interband
excitation~\cite{Silov2004} and by Faraday~\cite{Ganichev2006} or
Kerr effect~\cite{Yang2006}. In Faraday and Kerr effects the
rotation of polarization plane is proportional to the spin component
along the probe beam. This can complicate the measurement for
structure that generates only spin which is normal to the direction
of light propagation. However, an applied weak magnetic field
rotates the spin that leads to non-zero Faraday (Kerr)
signal~\cite{Kato2004a}.

The strong CISP dependence on the crystal axis along which the
electric field is applied was detected for [110]-grown QW in
Ref.~\cite{Sih2005}, reflecting the anisotropy of the SOC (SOC
contain the term proportional to $\sigma_zk_x$ with
$x||[\bar{1}10]$).

In conclusion, we present here the short overview of theoretical and
experimental works concerning CISP-effect and some related
phenomena. The different mechanisms leading to CISP in
semiconductors of various dimensionality are discussed, including
specificity of band structure and relaxation features. Our recent
results concerning the CISP in quasi-1D systems are presented as
well.

\section*{Acknowledgement}

We are grateful to L.E. Golub for useful discussions. This work was
supported by the Government of the Russian Federation (project No.
14.Z50.31.0021 with leading scientist M. Bayer).

\end{document}